\documentclass{article}
\usepackage{amssymb}

\input{tcilatex}
\begin{document}

\title{\textbf{Tunneling in attosecond optical ionization and a dynamical
time operator}}
\author{M. Bauer \\
Instituto de F\'{\i}sica\\
Universidad Nacional Aut\'{o}noma de M\'{e}xico\\
bauer@fisica.unam.mx}
\maketitle

\begin{abstract}
The conundrum \textit{parameter-operator} of time in quantum mechanics (QM),
as well as the \textit{time-energy uncertainty relation} and the \textit{%
tunneling} \textit{delay time,} have recently been addressed in attosecond
optical ionization experiments. The parameter status of time in the time
dependent Schr\"{o}dinger equation (TDSE) is supported by the well-known
Pauli's objection as well as by its interpretation as an emerging property
of entanglement with a classical environment. On the other hand, the
introduction of a self-adjoint dynamical time operator in Dirac's
formulation of electron's relativistic quantum mechanics (RQM), yields an
additional system observable that represents an internal time. In the
present paper the relation of this internal time with the parametric
(laboratory) time and its relevance to the tunneling measurements in these
experiments is examined within the standard framework of RQM.


\end{abstract}

\section{Introduction}

The conundrum \textit{parameter-operator} of time in quantum mechanics (QM),
as well as the \textit{time-energy uncertainty relation} and the \textit{%
tunneling} \textit{delay time,} have recently been addressed again in the
development of attosecond optical ionization experiments\cite{Eckle,Landsman}%
. The tunneling phenomenon, one of the earliest theoretical successes of QM,
has been extensively debated\ in relation to the question of the time the
particle spends in the barrier region. This has given rise to alternative
definitions of tunneling times but has not been definitively resolved\cite%
{Orlando,Landauer}. On the other hand, the technical development of
attosecond pulses of extreme ultaviolet radiation has allowed photionization
processes where a tunneling delay time can be measured and compared to
theoretical predictions, although using a time-energy uncertainty relation
associated with the commutation relation rightfully objected by Pauli\cite%
{Orlando,Maquet,Kullie2,Kullie}.

Indeed the existence of a time-energy uncertainty relation analogue to the
position-momentum one, conjectured by Heisenberg early on, faced from the
start Pauli's objection to the existence of a time operator, to quote\cite[%
p.63]{Pauli}: "...from the C.R. written above (cf. $[t,H]=i\hslash $) it
follows that $H$ possesses continously all eigenvalues from $-\infty $ to $%
+\infty $, whereas on the other hand, discrete eigenvalues of $H$ can be
present. \textit{We, therefore, conclude that the introduction of an
operator }$\mathit{t}$\textit{\ is basically forbidden and the time }$t$%
\textit{\ must necessarily be considered as on ordinary number ("c" number)
in Quantum Mechanics"}. In the time dependent Schr\"{o}dinger equation
(TDSE) time appears as a parameter, not an operator\cite{Pauli,Dirac}. This
led to a variety of alternative proposals for a time-energy uncertainty
relation and an extensive discussion of time in quantum mechanics throughout
several decades\cite{Muga,Muga2,Bauer1,Bush,Briggs,Hilgevoord}. Pauli's
argument, sustained also by the fact that the system's stability requires
the energy to have a finite minimum, is still subject of current research,
as well as the existence and meaning of a time-energy uncertainty relation%
\cite{Galapon,Boykin}. The undisputed experimental corroboration of Schr\"{o}%
dinger's equation supports the interpretation of the parameter $t$ as the
laboratory time. Its presence in the dynamical evolution of microscopical
systems (TDSE) has been atributed to the entanglement of these systems with
a macroscopic classical environment\cite{Briggs2}.

Recently however, it has been shown that Dirac's formulation of electron's
relativistic quantum mechanics (RQM) does allow the introduction of a
dynamical time operator that is self-adjoint\cite{Bauer}. Consequently, it
can be considered an additional system observable representing an internal
time, and proven to be subject to an uncertainty relation that circumvents
Pauli's objection. In the present paper it is shown that it provides an
equal footing of time and space in the analysis of the attosecond optical
ionization processes, as suggested in Ref.6\footnote{%
In this respect, Dodonov's\ quoted paper (Ref.4 in Ref. 6) , that claims
that no unambigous and generally accepted results have been obtained so far,
refers only to the present author paper of 1983 (Ann.Phys. \textbf{150, }1%
\textbf{\ } (1983) ), but not to the 2014 paper\cite{Bauer} that introduces
the dynamical time operator.}. These aspects are examined within the
standard framework of RQM. The definition and main properties of the
proposed time operator are recalled in Section 2. In Section 3 the ensuing
time-energy uncertainty relation is derived. It is also compared in Appendix
A to the Mandelstam-Tamm formulation extensively addressed in the discussion
of tunneling. Section 4 develops its application to the attosecond optical
ionization processes. Section 5 advances conclusions and possible
developments.

\section{The dynamical time operator in RQM}

A dynamical self-adjoint "time operator"%
\begin{equation}
\hat{T}=\mathbf{\alpha .\hat{r}/}c\mathbf{+}\beta \tau _{0}
\end{equation}%
has been introduced\cite{Bauer} in analogy to the Dirac free particle
Hamiltonian $\hat{H}_{D}=c\mathbf{\alpha .\hat{p}}+\beta m_{0}c^{2\text{ }},$
where $\alpha _{i}(i=1,2,3)$ and $\beta $ are the $4\times 4$ Dirac
matrices, satisfying the anticonmutation relations:%
\begin{equation}
\alpha _{i}\alpha _{j}+\alpha _{j}\alpha _{i}=2\delta _{ij}\ \ \ \ \ \alpha
_{i}\beta +\beta \alpha _{i}=0\ \ \ \ \beta ^{2}=1
\end{equation}%
$\tau _{0}$ represents in principle an internal property of the sysytem, to
be determined. In the Heisenberg picture, using the relations\cite%
{Thaller,Greiner,Messiah}:%
\begin{equation}
\mathbf{\alpha (}t\mathbf{)}=\mathbf{\alpha (}0\mathbf{)+\{\alpha (}0\mathbf{%
)-}c\mathbf{\hat{p}}/\hat{H}_{D}\}\{\exp (-2i\hat{H}_{D}t/\hslash )-1\}
\end{equation}%
\begin{equation}
\beta (t)=\beta (0)+\{\beta (0)-m_{0}c^{2}/\hat{H}_{D}\}\{\exp (-2i\hat{H}%
_{D}t/\hslash )-1\}
\end{equation}%
\begin{equation}
\mathbf{\hat{r}(}t\mathbf{)=\hat{r}(}0\mathbf{)+(}c^{2}\mathbf{\hat{p}}/\hat{%
H}_{D})t+i(c\hslash /2\hat{H}_{D})\{\exp (-2i\hat{H}_{D}t/\hslash )-1\}
\end{equation}%
the time evolution of the time operator is given by:%
\begin{eqnarray}
\hat{T}(t) &=&\mathbf{\alpha (}t\mathbf{).\hat{r}(}t\mathbf{)/}c\mathbf{+}%
\beta (t)\tau _{0}=  \nonumber \\
&=&\mathbf{\alpha (}0\mathbf{).\hat{r}(}0\mathbf{)/}c+\beta (0)\tau _{0}+%
\mathbf{\alpha (}0\mathbf{).(}c^{2}\mathbf{\hat{p}}/\hat{H}%
_{D})t+oscillating\ terms  \nonumber \\
&=&\hat{T}(0)+(c\mathbf{\hat{p}}/\hat{H}_{D})^{2}t+oscillating\ terms
\end{eqnarray}%
where use has been made of: 
\begin{equation}
c\mathbf{\alpha (}0\mathbf{).(}c\mathbf{\hat{p}}/\hat{H}_{D})=\left[ \frac{d%
\mathbf{\hat{r}}}{dt}\right] _{t=0}.\mathbf{(}c\mathbf{\hat{p}}/\hat{H}_{D})=%
\mathbf{(}c\mathbf{\hat{p}}/\hat{H}_{D})^{2}+oscillating\ terms
\end{equation}%
Thus $\hat{T}(t)$\ exhibits a linear dependence on $t$\ with a superimposed
oscillation (Zitterbewegung), as occurs with the time development of the
position operator $\mathbf{\hat{r}}(t)$.

In this formulation, $\tau _{0}$ plays the role of an invariant quantity in
the $(\boldsymbol{r},\tau )$ space, i.e., $\tau _{0}^{2}=\tau ^{2}-(%
\boldsymbol{r}/c)^{2}$, as $m_{0}c^{2}$ plays in the $(\boldsymbol{p},E)$
space, namely $(m_{0}c^{2})^{2}=E^{2}-(c\boldsymbol{p})^{2}$. To mantain the
fundamental indeterminacy modulo $n2\pi $\ ($n$ an integer) in the phase of
the complex eigenfunctions one has to set, for $n=1$\textit{:}%
\begin{equation}
\tau _{0}=2\pi \hbar /<\beta >\varepsilon =h/m_{0}c^{2}
\end{equation}%
This is the de Broglie period\cite{Broglie,Bayliss}. Together with the
Compton wave length, it sets a unified spacetime Compton scale that limits
the wave packets width in space and time before negative energy and negative
time components (particle and antiparticle) occur significantly. Moreover,
it supports the existence of an internal property, the de Broglie clock with
a period $\tau _{0}=h/m_{0}c^{2}$\cite{Ferber,Lan,Catillon}.

It is now important to note that in the non relativistic energies case $%
\left\langle \hat{H}_{D}\right\rangle \simeq m_{0}c^{2}$ , one has,
neglecting oscillating terms in Eq.6:%
\begin{equation}
\left\langle \hat{T}(t_{2})\right\rangle -\left\langle \hat{T}%
(t_{1})\right\rangle \simeq (v_{gp}/c)^{2}(t_{2}-t_{1})
\end{equation}%
\textit{Thus dynamical (internal) intervals are contracted with respect to
parametric (external) intervals.}

On the other hand, in the case of ultra relativistic energies, $\left\langle 
\hat{H}_{D}\right\rangle \simeq cp$ and Eq.6 yields: 
\begin{equation}
\left\langle \hat{T}(t_{2})\right\rangle -\left\langle \hat{T}%
(t_{1})\right\rangle \simeq \left\langle (cp/cp)^{2}\right\rangle
(t_{2}-t_{1})=t_{2}-t_{1}
\end{equation}%
\textit{Dynamical (internal) intervals coincide with parametric (external)
intervals.}

Finnaly, the time operator, being self-adjoint, is the generator of
continous momentum displacements, and thus indirectly of continous energy
dispalcements within the positive and the negative energy branches, but not
across the energy gap. In this way Pauli's objection is circumvented.

\section{The time-energy uncertainty relation}

The time operator and the Dirac Hamiltonian satisfy the commutaion relation%
\cite{Bauer}:%
\begin{equation}
\lbrack \hat{T},\hat{H}_{D}]=i\hbar \{I+2\beta K\}+2\beta \{\tau _{0}\hat{H}%
_{D}-m_{0}c^{2}\hat{T}\}
\end{equation}%
where $K=\beta (2\mathbf{s.l}/\hbar ^{2}+1)$\ is a constant of motion\cite%
{Thaller}. In the usual manner an uncertainty relation follows, namely:%
\begin{equation}
(\Delta T)(\Delta H_{D})\geq (\hbar /2)\left\vert 1+2<\beta K>\right\vert
=(3\hbar /2)\left\vert 1+\frac{4}{3}\left\langle \mathbf{s.l}/\hbar
^{2}\right\rangle \right\vert
\end{equation}%
where \ $\Delta T=\sqrt{\left\langle \hat{T}^{2}\right\rangle -\left\langle 
\hat{T}\right\rangle ^{2}}$and \ $\Delta H=\sqrt{\left\langle \hat{H}%
_{D}{}^{2}\right\rangle -\left\langle \hat{H}_{D}\right\rangle ^{2}\text{.}}$

To be noted is that the uncertainty of the present time operator is related
to the uncertainty in position $\Delta \mathbf{r}$ , in the same way as the
energy uncertainty is related to the momentum uncertainty $\Delta \mathbf{p.}
$ Indeed:%
\begin{eqnarray*}
(\Delta T)^{2} &=&\left\langle \hat{T}^{2}\right\rangle -\left\langle \hat{T}%
\right\rangle ^{2}=\left\langle \mathbf{\hat{r}}^{2}/c^{2}+\tau
_{0}^{2}\right\rangle -\left\langle \hat{T}\right\rangle ^{2}= \\
&=&\{(\Delta \mathbf{r})^{2}+\left\langle \mathbf{\hat{r}}\right\rangle
^{2}\}/c^{2}+\tau _{0}^{2}-\{\left\langle \mathbf{\alpha .\hat{r})/}c\mathbf{%
+}\beta \tau _{0}\right\rangle ^{2}\}= \\
&=&(\Delta \mathbf{r})^{2}/c^{2}+\tau _{0}^{2}(1-\left\langle \beta
\right\rangle ^{2})+(\left\langle \mathbf{\hat{r}}\right\rangle
^{2}-\left\langle \mathbf{\alpha .\hat{r}}\right\rangle ^{2})/c^{2}-2\tau
_{0}\left\langle \mathbf{\alpha .\hat{r})/}c\right\rangle \left\langle \beta
\right\rangle
\end{eqnarray*}%
Thus:%
\begin{equation}
\Delta T\gtrapprox \Delta \mathbf{r}/c
\end{equation}%
and similarly: 
\begin{eqnarray}
\ (\Delta H_{D})^{2} &=&\left\langle \hat{H}_{D}^{2}\right\rangle
-\left\langle \hat{H}_{D}\right\rangle ^{2}=c^{2}\left\langle \mathbf{\hat{p}%
}^{2}\right\rangle +(m_{0}c^{2})^{2}-\left\langle \hat{H}_{D}\right\rangle
^{2}  \nonumber \\
&=&c^{2}\{(\Delta \mathbf{p})^{2}+\left\langle \mathbf{\hat{p}}\right\rangle
^{2}\}+(m_{0}c^{2})^{2}-\left\langle \hat{H}_{D}\right\rangle ^{2}\gtrapprox
c^{2}(\Delta \mathbf{p})^{2}
\end{eqnarray}%
Then%
\begin{equation}
(\Delta T)(\Delta H_{D})\gtrapprox (\Delta \mathbf{r)(}\Delta \mathbf{p)\geq 
}(3\hbar /2)
\end{equation}%
The association of \ $\Delta T$\ \ with \ $\Delta \mathbf{r}$\ , and of \ $%
\Delta H_{D}$\ \ with \ $\Delta \mathbf{p}$\ ,\ corresponds to Bohr's
interpretation: the width of a wave packet, complementary to its momentum
dispersion and thus to its energy dispersion, measures the uncertainty in
the time of passage at a point of the trajectory.

In the presence of potentials dependent only on position, e.g., Coulomb type
potentials, the above result is maintained as: 
\begin{equation}
\lbrack \hat{T},\hat{H}_{D}+V(\mathbf{\hat{r}})]=[\hat{T},\hat{H}_{D}]
\end{equation}%
and the same uncertainty relation will follow.If in addition there is
spherical symmetry, the initial position and momentum expectation values
vanish, i.e. $\left\langle \mathbf{r}\right\rangle =0$ and $\left\langle 
\mathbf{p}\right\rangle =0$\ . Then Eqs.13\ and 14\ become:

\[
(\Delta T)^{2}=(\Delta \mathbf{r})^{2}/c^{2}+\tau _{0}^{2}(1-\left\langle
\beta \right\rangle ^{2})\geqq (\Delta \mathbf{r})^{2}/c^{2} 
\]

\[
\ (\Delta H_{D})^{2}=c^{2}(\Delta \mathbf{p})^{2}+(m_{0}c^{2})^{2}(1-\left%
\langle \beta \right\rangle ^{2})\geqq c^{2}(\Delta \mathbf{p})^{2} 
\]

\section{Tunneling time in attosecond optical ionization}

The sudden onset of a laser pulse opens the electron bound state at energy $%
E_{0}=-I_{p}$ to tunneling through a barrier created by an effective
potential in the direction of the pulse polarization, modelled as\cite%
{Orlando,Kullie,Kullie2}:%
\begin{equation}
V_{eff}=-\frac{Z_{eff}e}{\left\vert \mathbf{r}\right\vert }-\mathbf{F.r}
\end{equation}%
The first term is the binding Coulomb potential and the second is the dipole
interaction with a pulse of maximum intensity $F$. The barrier width $d_{B}$%
\ in the radial direction, say $x$,of the electric field, is given by the
difference between the entrance $x_{e,-}$\ and exit $x_{e,+}$ points of the
barrier (Fig.1 of Ref. \ ), i.e., the solutions to the equation:%
\[
-\frac{Z_{eff}e}{x}-Fx=-I_{p} 
\]%
yielding:%
\begin{equation}
d_{B}(F)\doteq \{x_{e,+}-x_{e,-}\}=(I_{p}/F)\sqrt{1-4Z_{eff}eF/I_{p}^{2}}
\end{equation}%
as given in Eq.13 of Ref.6.

As shown in Section 3 above, the minimum time uncertainty for spherical
symmetry is:%
\begin{equation}
\Delta T=\Delta \mathbf{r}/c=\left\langle r^{2}\right\rangle ^{1/2}/c
\end{equation}%
where integration is carried over all directions. If now one assumes that
the tunnelig internal time $\bar{\tau}_{T}$ in one direction is equal to $%
(1/4\pi )\Delta T$ and that the time uncertainty is of the order of the time
uncertainty associated with the barrier width $(\delta r=d_{B}(F))$, one
concludes that for a single direction the internal tunneling time is given
by:%
\begin{equation}
\bar{\tau}_{T}\approx (1/4\pi )d_{B}(F)/c
\end{equation}%
i.e., $\bar{\tau}_{T}$ is proportional to the time it would take a photon to
traverse the barrier width. Then from Eq.17, the laboratory tunneling time
in the non relativistic regime is given by:%
\begin{equation}
\Upsilon _{T}\approx \bar{\tau}_{T}/(v_{gp}/c)^{2}\approx \{(1/4\pi
)d_{B}(F)/c\}/(v_{gp}/c)^{2}\gg \bar{\tau}_{T}
\end{equation}%
There is thus a linear relation between laboratory tunneling time and
barrier width, as has been experimentally obtained (Fig.3(d) of Ref.2).

The enhancing factor between internal and (laboratory) parameter times can
be evaluated as follows. Ref.2 reports an electron tunneling time of $40\ as$
for a barrier width of $13\ a.u.=6.88\ \mathring{A}$ . This gives a
tunneling velocity $v_{gp}=\frac{6.88}{40}\times 10^{10}$ $cm/s$ , and a
ratio $1/(v_{gp}/c)^{2}=304.22$. . It follows then:%
\[
\Upsilon _{T}\approx (1/4\pi )304.22[d_{B}(F)/c] 
\]%
For a barrier width of $20\ a.u.=10.58\ \mathring{A}$ one obtains:%
\begin{equation}
\Upsilon _{T}\approx (1/4\pi )304.22[d_{B}(F)/c]=24.22\times (10.58/3)\
as=85.4\ as
\end{equation}%
while for a barrier width of $8\ a.u.=4.233\ \mathring{A}$ one obtains:%
\begin{equation}
\Upsilon _{T}\approx (1/4\pi )304.22[d_{B}(F)/c]=24.22\times (4.233/3)\
as=34.2\ as
\end{equation}%
These value compare well with the experimental results shown in Fig.3(d) of
Ref.2.\ The straight line joining these values has a slightly different
slope of that of the FPI (Feynman Path Integral) quantum mechanical result,
but falls within the experimental uncertainties.

The dependence on the field intensity is obtained using Eq.18, namely:%
\begin{equation}
\Upsilon _{T}\approx (1/4\pi )[(I_{p}/F)\sqrt{1-4Z_{eff}eF/I_{p}^{2}}%
]/c(v_{gp}/c)^{2}
\end{equation}%
which gives the observed shape of the dependence of the barrier width on the
field intensity (Fig.3(b) of Ref.2).

\section{Conclusion}

The dynamical time operator provides a straightforward explanation within
standard RQM of the tunneling times measured in the photoionization
experiments. As an observable, it introduces an internal time in addition to
the parameter (laboratory) time in the TDSE that has been shown to be an
emergent property arising from the entanglement of a microscopic system with
a classical environment in an overall closed time independent system, this
property being apparent only to an internal observer\cite{Briggs2,Moreva}.
There is no conumdrum \textit{parameter-operator} of time in quantum
mechanics, as both times are seen to play a role in RQM. Also predicted is
an enhancement at low energies between internal and laboratory tunneling
times that fits the measurements in attosecond optical ionization
experiments.

Based on the position observable, the time operator is expected to exhibit a
Zitterbewegung behaviour about its linear dependence on $t$. As occurs with
the position one, its observation is beyond current technical possibilities.
However it may be observable in systems that simulate Dirac's Hamiltonian,
where position Zitterbewegung has allready been exhibited experimentally\cite%
{Cserti,Gerritsma,LeBlanc}. A corresponding time operator can be constructed
in each case and perhaps its properties may be exhibited in similar
experiments.

Finally, general relativity accords a dynamical behaviour to space-time,
firmly confirmed recently by the detection of gravitational waves. As a
dynamical time is definitively incompatible with a time parameter, this
becomes from the start a fundamental "problem of time" in quantum gravity%
\cite{Anderson,Isham,Butterfield}. Whether the time operator here introduced
has a relevance in this subject, is a venue to be considered\cite{Bauer3}.

\section{Appendix A: Mandelstam-Tamm time-energy uncertainty relation}

As an observable, the time operator can be subject to the Mandelstam-Tamm
(MT) formulation of a time-energy uncertainty relation within standard QM%
\cite[p.319]{Messiah}, to wit: any observable $A$ represented by a
self-adjoint operator $\hat{A}$ not explicitly dependent on time, satisfies
the dynamical equation:

\begin{equation}
(i\hbar )\frac{d}{dt}<\hat{A}>=<[\hat{A},\hat{H}]>
\end{equation}%
From the commutator\ $[\hat{A},\hat{H}]$\ it follows that the uncertainties
defined $\Delta \hat{A}$ and $\Delta \hat{H}$ satisfy the relation: 
\begin{equation}
(\Delta \hat{A})(\Delta \hat{H})\geq (1/2)\mid <[\hat{A},\hat{H}]>\mid
=(1/2)\left\vert \frac{d}{dt}<\hat{A}>\right\vert
\end{equation}%
Then, associated to any system observable $\hat{A}$ , a related time
uncertainty\textbf{\ }is\textbf{\ }defined as:%
\begin{equation}
\Delta \hat{T}_{\hat{A}}^{mt}=\frac{\Delta \hat{A}}{\mid \frac{d}{dt}<\hat{A}%
>\mid }
\end{equation}%
From Eqs. 17 and 18, it then follows that:%
\begin{equation}
(\Delta \hat{T}_{\hat{A}}^{mt})(\Delta \hat{H})\geq (\hbar /2)
\end{equation}%
This is the Mandelstam-Tamm time-energy uncertainty relation. $\Delta \hat{T}%
_{\hat{A}}^{mt}$ can be interpreted as \textit{"the time required for the
center }$\left\langle \hat{O}\right\rangle $\textit{\ of this distribution
to be displaced by an amount equal to its width }$\Delta \hat{A}$\textit{"}%
\cite{Messiah}.

Now let $\hat{A}$ be the dynamical time operator $\hat{T}=(\mathbf{\alpha .r}%
)/c+\beta \tau _{0}$. Then, from Eq. 20 and Eq.12, one obtains:

\begin{equation}
\Delta T_{\hat{T}}^{mt}\approx \frac{\Delta \hat{T}}{\left\vert \left\langle
I+2\beta K\right\rangle \right\vert }
\end{equation}%
It follows that:%
\begin{equation}
\frac{\Delta \hat{T}}{\left\vert \left\langle I+2\beta K\right\rangle
\right\vert }(\Delta \hat{H}_{D})\geq (\hbar /2)
\end{equation}%
or%
\begin{eqnarray}
(\Delta \hat{T})(\Delta \hat{H}_{D}) &\geq &(\hbar /2)\left\vert
\left\langle I+2\beta K\right\rangle \right\vert =(3\hbar /2)\left\vert 1+%
\frac{4}{3}\left\langle \mathbf{s.l}/\hbar ^{2}\right\rangle \right\vert = 
\nonumber \\
&=&(3\hbar /2)\left\vert 1+\frac{2}{3}\left\langle (\mathbf{j}^{2}-\mathbf{l}%
^{2}-\mathbf{s}^{2})/\hbar ^{2}\right\rangle \right\vert \geq 3\hbar /2)
\end{eqnarray}

In the non relativistic limit $\left\langle \hat{H}_{D}\right\rangle \simeq
m_{0}c^{2}$, neglecting the oscillating terms, Eq.3\ yields:

\begin{equation}
\left\langle \hat{T}(t)\right\rangle \simeq \left\langle \hat{T}%
(0)\right\rangle +\left\langle (cp/m_{0}c^{2})^{2}\right\rangle
t+...=\left\langle \hat{T}(0)\right\rangle +(v_{gp}/c)^{2}t
\end{equation}%
Thus:%
\[
\frac{d<\hat{T}>}{dt}=\left\langle (cp/m_{0}c^{2})^{2}\right\rangle
=(v_{gp}/c)^{2} 
\]%
and%
\begin{equation}
\Delta T_{\hat{T}}^{mt}\simeq \frac{\Delta \hat{T}}{(v_{gp}/c)^{2}}\gg
\Delta \hat{T}
\end{equation}%
as $v_{gp}<<c$. The Mandelstam-Tamm uncertainty associated with the
observable $\hat{T}$ overestimates largely the internal time standard
uncertainty.


\begin{thebibliography}{99}
\bibitem{Eckle} Eckle, P. etal, "Attosecond Ionization and Tunneling Delay
Time Measurements in Helium", Science \textbf{322}, 1525 (2008)

\bibitem{Landsman} Landsman, A.S.\textit{\ et al}., "Ultrafast resolution of
tunneling delay time", Optica \textbf{1}, 343 (2014)

\bibitem{Orlando} Orlando, G. \textit{et al.}, "Tunneling time, what does it
mean?", J. Phys. B:At. Mol. Opt. Phys. \textbf{47}, 204002 (2014)

\bibitem{Landauer} Landauer, r. and Ph. Martin, "Barrier interaction time in
tunneling", Rev.Mod.Phys. \textbf{66.} 217-228 (1994)

\bibitem{Maquet} Maquet, M., J. Caillat and R. Ta\"{\i}eb, "Attosecond
delays in photoionization: time\textit{\ and} quantum mechanics", J.Phys. B:
At.Mol.Opt.Phys. \textbf{47}, 204004 (2014)

\bibitem{Kullie2} Kullie, O., "Tunneling time in attosecond experiments and
the time-energy uncertainty relation", Phys.Rev. A\textbf{\ 92}, 052118
(2015)

\bibitem{Kullie} Kullie, O., "Tunneling time in attosecond experiments,
intinsic-type of time. Keldysh amd Mandelstam-Tamm time", J.Phys. B:
At.Mol.Opt.Phys. \textbf{49}, 095601 (2015)

\bibitem{Pauli} Pauli, W., \textit{"The general principles of quantum
mechanics"}, Springer-Verlag, Berlin Heidelberg, footnote p.63 (1980)

\bibitem{Dirac} Dirac, P.A.M., \textit{\textquotedblleft The principles of
quantum mechanics"} (4th ed.), Oxford, Clarendon Press (1958)

\bibitem{Muga} Muga, J.G., R. Sala Mayato and I.L. Egusquiza ,
"Introduction", in J.G. Muga, R. Sala Mayato, I.L. Egusquiza (eds.) "\textit{%
Time in Quantum Mechanics"}, pp 1-28 Berlin Springer (2002); reprinted as "%
\textit{Time in Quantum Mechanics}, \textit{Vol. 1"}, Lect. Notes Phys. 
\textbf{734}, Springer-Verlag, Berlin (2008)

\bibitem{Muga2} Muga, J.G., A. Ruschhaupt and A. del Campo (eds), "\textit{%
Time in Quantum Mechanics}, \textit{Vol. 2"}, Lect. Notes Phys. \textbf{789}%
, Springer-Verlag, Berlin (2009).

\bibitem{Bauer1} Bauer, M. and P.A. Mello, "The time-energy uncertainty
relation", Ann.Phys. \textbf{111}, 38-60 (1978)

\bibitem{Bush} Busch, P., "The time-energy uncertainty relation", chapter 3
in J.G. Muga, R. Sala Mayato, I.L. Egusquiza (eds.) "\textit{Time in Quantum
Mechanics"}, Berlin Springer (2002); revised version:
arXiv:quant-ph/0105049v3 (2007)

\bibitem{Briggs} Briggs, J., "A derivation of the time-energy uncertainty
relation", Journal of Physics: Conference Series \textbf{99}, 012002 (2008)

\bibitem{Hilgevoord} Hilgevoord, J., "Time in Quantum Mechanics", Am.J.Phys.%
\textbf{\ 70}, 301-306 (2002)

\bibitem{Galapon} Galapon, E.A., "Post-Pauli's Theorem Emerging Perspective
on Time In Quantum Mechanics", Chapter 3 in\ Lect. Notes Phys. \textbf{789},
25-63 (2009)

\bibitem{Boykin} Boykin, T.B., N. Kharche and G. Klimeck, "Evolution time
and energy uncertainty", Eur.J.Phys. \textbf{28}, 673-678 (2007)

\bibitem{Briggs2} Briggs, J.S. and Jan M. Rost, \textquotedblleft Time
dependence in quantum mechanics\textquotedblright , Eur.Phys.J. \textbf{10},
(2000); \textquotedblleft On the Derivation of the Time-dependent Equation
of Schr\"{o}dinger\textquotedblright , Found.Phys. \textbf{31}, (2001)

\bibitem{Bauer} Bauer, M., \textquotedblleft A dynamical time operator in
Dirac's relativistic quantum mechanics", Int.J.Mod.Phys. A \textbf{29},
1450036 (2014)

\bibitem{Thaller} Thaller, B., \textit{\textquotedblleft The Dirac Equation"}%
, Springer-Velag, Berlin Heidelberg New York (1992)

\bibitem{Greiner} Greiner, W., \textit{\textquotedblleft Relativistic
Quantum Mechanics - Wave equations"}, (3$^{\text{d}}$ ed.) Springer, Berlin
Heidelberg New York (2000)

\bibitem{Messiah} Messiah, A., \textit{"Quantum Mechanics"}, Vol.I, p. 442,
North-Holland Publishing Company, Amsterdam, and John Wiley\&Sons, New York
London Sidney, 4th printing (1966)

\bibitem{Broglie} de Broglie, L., Ph.D. thesis; Ann. Phys.,Ser. 10$^{\text{e}%
}$, t. III (1925). English translation reprinted in Ann.Fond.Louis de
Broglie \textbf{17}, 92 (1992)

\bibitem{Bayliss} Baylis, W.E., \textquotedblleft De Broglie waves as an
effect of clock desynchronization", Can.J.Phys. \textbf{85}, 1317-1323 (2007)

\bibitem{Ferber} Ferber, R., "A Missing Link: What is behind de Broglie's
"periodic phenomenon"?, Found.Phys.Lett. \textbf{9}, 575-586 (1996)

\bibitem{Lan} Lan, S.Y. et al., "A Clock Directly Linking Time to a Particle
Mass", Science \textbf{339}, 554-557 (2013)

\bibitem{Catillon} Catillon, P. \textit{et al.}, \textquotedblleft A Search
for the de Broglie Particle Internal Clock by means of Electron
Channeling\textquotedblright , Found.Phys. \textbf{38}, 659-664 (2008)

\bibitem{Moreva} Moreva, E. \textit{et al.}, "Time from quantum
entanglement: an experimental illustration", arXiv:1310.4691v1 [quant-ph]
(2013); "The time as an emergent property of quantum mechanics, a synthetic
description of a first experimental approach", J.Phys.: Conference Series 
\textbf{626}, 012019 (2015)

\bibitem{Cserti} Cserti, J. and G. David, "Unified description of
Zitterbewegung for spintronic, graphene, and superconducting systems",
Phys.Rev. B \textbf{74}, 172305 (2006)

\bibitem{Gerritsma} Gerritsma, R. \textit{et al}., \textquotedblleft Quantum
simulation of the Dirac equation", Nature \textbf{463}, 68-71 (2010)

\bibitem{LeBlanc} LeBlanc, L.J. \textit{et al}., \textquotedblleft Direct
observation of Zitterbewegung in a Bose-Einstein condensate", New J.Phys. 
\textbf{15}, 073011 (2013)

\bibitem{Anderson} Anderson, E., "The problem of time in quantum gravity",
Ann.Phys. (Berlin) \textbf{524}, 757-786 (2012) and references therein; also
arXiv:1009.2157v3 [gr-qc]

\bibitem{Isham} Isham, C.J., "Canonical Quantum Gravity and the Problem of
Time", arXiv:gr-qc/9210011v1 (1992); "Prima Facie Questions in Quantum
Gravity", arXiv:gr-qc/9310031v1 (1993)

\bibitem{Butterfield} Butterfield, J. and C.J. Isham, "On the Emergence of
Time in Quantum Gravity", arXiv:gr-qc/9901024v1 (1999)

\bibitem{Bauer3} Bauer, M., "Quantum Gravity and a Time Operator in
Relativistic Quantum Mechanics", arXiv:gr-qc/1605.01659 (2016)
\end{thebibliography}
\end{document}